\documentclass[fleqn,10pt]{wlscirep}
\usepackage[utf8]{inputenc}
\usepackage[T1]{fontenc}
\title{Tumor Location-weighted MRI-Report Contrastive Learning: A Framework for Improving the Explainability of Pediatric Brain Tumor Diagnosis}

\author[1,2,3]{Sara Ketabi}
\author[6,7]{Matthias W. Wagner}
\author[9]{Cynthia Hawkins}
\author[8]{Uri Tabori}
\author[1,4,6,10]{Birgit Betina Ertl-Wagner}
\author[1,2,3,4,5,6,10*]{Farzad Khalvati}
\affil[1]{Neurosciences and Mental Health Research Program, The Hospital for Sick Children, Toronto, Canada}
\affil[2]{Department of Mechanical
and Industrial Engineering, University of Toronto, Toronto, Canada}
\affil[3]{Vector Institute for Artificial Intelligence, Toronto, Canada}
\affil[4]{Institute of
Medical Science, University of Toronto, Toronto, Canada}
\affil[5]{Department of Computer Science, University of Toronto, Toronto, Canada}
\affil[6]{Department of Diagnostic and Interventional Radiology,
The Hospital for Sick Children, Toronto, Canada}
\affil[7]{Department of Diagnostic and Interventional Neuroradiology,
University Hospital Augsburg, Augsburg, Germany}
\affil[8]{Division of Hematology and Oncology, The Hospital for Sick
Children, Toronto, Canada}
\affil[9]{Paediatric Laboratory Medicine, Division of Pathology, The
Hospital for Sick Children, Toronto, Canada}
\affil[10]{Department of Medical Imaging, University of Toronto, Toronto,
Canada}

\affil[*]{farzad.khalvati@utoronto.ca}

\begin{abstract}
Despite the promising performance of convolutional neural networks (CNNs) in brain tumor diagnosis from magnetic resonance imaging (MRI), their integration into the clinical workflow has been limited. That is mainly due to the fact that the features contributing to a model's prediction are unclear to radiologists and hence, clinically irrelevant, i.e., lack of explainability. As the invaluable sources of radiologists' knowledge and expertise, radiology reports can be integrated with MRI in a contrastive learning (CL) framework, enabling learning from image-report associations, to improve CNN explainability. In this work, we train a multimodal CL architecture on 3D brain MRI scans and radiology reports to learn informative MRI representations. Furthermore, we integrate tumor location, salient to several brain tumor analysis tasks, into this framework to improve its generalizability. We then apply the learnt image representations to improve explainability and performance of genetic marker classification of pediatric Low-grade Glioma, the most prevalent brain tumor in children, as a downstream task. Our results indicate a Dice score of 31.1\% between the model's attention maps and manual tumor segmentation (as an explainability measure) with test classification performance of 87.7\%, significantly outperforming the baselines. These enhancements can build trust in our model among radiologists, facilitating its integration into clinical practices for more efficient tumor diagnosis.
\end{abstract}
\begin{document}

\flushbottom
\maketitle
%
%
\thispagestyle{empty}

\noindent \textbf{Keywords:} Contrastive Representation Learning, MRI, Radiology Report, Explainability, Glioma

\section*{Introduction}

Convolutional neural networks (CNNs), as a subset of deep learning (DL), have yielded high performance in various brain tumor diagnosis tasks using magnetic resonance imaging (MRI) in a non-invasive manner \cite{namdar2024improving,tak2024noninvasive}. Nonetheless, these models are not widely trusted and verified by radiologists, hindering their adoption in clinical settings. This can have several reasons, one of the primary ones being \emph{lack of explainability}. Explainability refers to the ability of a DL model to focus on relevant parts of input data for making a certain prediction. In the case of medical image analysis, explainability makes a CNN model diagnose similarly to radiologists and attend to appropriate parts of an input image. Therefore, a potential approach for improving model explainability can be integrating sources of information derived from the knowledge of radiologists into the CNN's framework.

Different data sources have been utilized in the literature to enhance CNN explainability. These include semantic features  \cite{shen2019interpretable}, i.e., important imaging features annotated by radiologists corresponding to a certain disease, and eye-gaze information \cite{karargyris2021creation}, which is associated with parts of an image attended by a radiologist for making a diagnosis. However, these sources are usually expensive and labor-intensive, making them very challenging to obtain and incorporate into CNN models. Radiology reports, on the other hand, represent an invaluable data modality expressed in radiologists' language, which are easily accessible in most medical imaging datasets and can be used for enhancing model explainability. \emph{Contrastive learning (CL)}, as a subset of self-supervised learning (SSL), can utilize the correspondence between images and reports, without requiring any external labels, and serve as an effective method for leveraging radiology reports into CNN-based DL frameworks for improving CNN explainability.

Developing CL-based DL architectures on medical images and radiology reports has been investigated in several works for chest X-ray, which is a 2D imaging modality, based on the global and local correlation between images and reports \cite{huang2021gloria,wang2022multi,ji2021improving}. Global correlation corresponds to the association between global image and report representations, while local correlation refers to the alignment between local image patches and report words. For 3D imaging, CL has been used to align the global representations of brain MR images and radiology reports \cite{lei2023unibrain}. However, finding the local correlation between the two data modalities, which can be an effective way of improving the explainability of imaging-based models, should also be explored. Moreover, although adjusting the representations based on relevant additional variables has been studied in image-based SSL frameworks \cite{dufumier2021contrastive}, the incorporation of such variables into multimodal CL frameworks in order to improve the representation learning task has yet to be investigated, given the heterogeneous nature of clinical data.

In this work, we propose a deep CL framework on a set of MRI and report pairs to learn useful image representations which are semantically close to the corresponding text representations. Furthermore, we leverage tumor location, as a determining discrete variable in different brain tumor diagnostic tasks, into the CL framework to enhance the representation learning. Finally, we apply the learnt image representations to classify the genetic markers of pediatric Low-grade Glioma (pLGG), as a downstream task, to improve both explainability and performance.

PLGG is one of the most common brain tumors in children, with a prevalence rate of 30-40\%. Identifying the genetic markers or molecular subtypes of pLGG is highly important for tumor prognosis, risk stratification, and effective targeted treatment planning \cite{fangusaro2024pediatric,trasolini2022mr}. The major clinical method for identifying pLGG genetic markers is through biopsy, which is associated with several pitfalls, including being invasive and difficulty in accessing the tumor \cite{van2020pitfalls}. Enhancing the explainability of CNNs in detecting these genetic markers accelerates model adoption in the clinical workflow as a non-invasive diagnostic tool.

\subsection*{Explainability Improvement in Medical Image Classification}

  Previous studies have utilized different sources of domain knowledge in model training to make the model focus on relevant parts of the image. Semantic features, i.e., important imaging features annotated by radiologists, were applied in \cite{shen2019interpretable} to train a 3D CNN framework on computed tomography images for predicting binarized low-level lung-related semantic features
along with  classification labels indicating the malignancy of a lung nodule. Eye-gaze information, which specifies image areas important to the radiologist for making a diagnosis, was used in \cite{karargyris2021creation} to train a U-Net model. The encoder part of this model performs chest X-ray classification, and the decoder aims to predict eye-gaze heatmaps. It was quantitatively shown that adding this decoder branch to the baseline classification improves the quality of the model's attention maps \cite{ketabi2023multimodal}. 

Despite the effect of such sources of expert knowledge on boosting DL explainability, they are typically labor-intensive and challenging to obtain in medical settings. However, radiology reports, describing radiologists' opinions and findings on medical images, can be readily accessible and efficiently integrated into image-based DL models to guide the model's attention to relevant image areas. 

\subsection*{Performance Improvement in Medical Image Classification via Contrastive Learning on Medical Images and Radiology Reports}

Training CL-based architectures to learn generalizable representations of 2D medical images and radiology reports has been investigated in several works in the literature. \cite{wang2022multi,huang2021gloria,ji2021improving} developed CL on chest X-ray and report texts for learning the global and local interactions between image and text representations. Applying the learnt image representations to downstream classification tasks, they demonstrated considerable performance improvements over image-based models trained from scratch or initialized with ImageNet weights.
 
However, very limited works have investigated CL to find the correlation between 3D medical images, which are more complex in composition than 2D images, and radiology reports. \cite{lei2023unibrain} developed a pretraining framework on paired brain MRI, radiology reports, and disease-related queries for aligning image-text global representations as well as finding the correlation between global image feature map and disease queries. Nonetheless, developing an appropriate module for finding local image-text correlations for 3D imaging has yet to be investigated. Furthermore, including  additional variables pertinent to datapoints, alongside medical images and radiology reports, can help in adjusting the distance between negative (mismatched) representations. 
Although this idea was proposed in \cite{dufumier2021contrastive} by incorporating patient age into an MRI-based CL framework, including external variables in multimodal image-text frameworks should also be explored.

To the best of our knowledge, this is the first study that develops an image-text CL framework for optimizing both global and local interactions between MRI, as a 3D medical imaging modality, and radiology reports and utilizes a discrete variable, i.e., tumor location, for regulating the distance between mismatched image and report representations. Furthermore, in addition to improving the classification, we explore the novel application of this framework and semantic correspondence between MRI and reports in improving the explainability of image-based models.

\noindent Our contributions in this work can be summarized into three main folds:
\begin{itemize}
    \item Training a deep MRI-report CL framework for learning generalizable MRI representations
    \item Incorporating tumor location into the CL framework to enhance the representation learning process
    \item Fine-tuning the CL-based MRI representations on pLGG genetic marker classification to improve the explainability and performance of this downstream task
\end{itemize}

\section*{Materials and Methods}

\subsection*{Data}
All methods of this retrospective study were performed in accordance with the guidelines and regulations of the research ethics board of The Hospital for Sick Children (SickKids) (Toronto, Canada), which approved the study and waived informed consent. In this study, we use two main datasets, both coming from SickKids. Table \ref{tab5} in Appendix shows the distribution of the ground-truth genetic marker labels extracted from biopsy results for each dataset. Dataset 1 corresponds to patients aged between 0 and 17 diagnosed with pLGG who underwent an MRI examination between 2000 and 2018. It contains 341 3D brain MR images (Fluid attenuated inversion recovery (FLAIR) sequence) from 341 unique patients, corresponding radiology reports, manual tumor segmentation masks, genetic marker labels, and tumor location information. The tumor location information spans three different classes specifying the location of the tumor in the brain, namely Supratentorial, Infratentorial, and Transtentorial. 
For the CL-based pretraining task, we only used the 341 MR images and radiology reports. 
For fine-tuning the pretrained MRI representation on the downstream classification task, we included 204 datapoints associated with the two most important genetic markers, i.e., BRAF Fusion and BRAF V600E Mutation, and excluded other 137 datapoints. Manual tumor segmentation masks were not used during model training by any means, applied only for evaluating the explainability of the MRI-based classification model. Both MR images and segmentation masks were preprocessed and resized to the shape of (240,240,155), following the methods described in \cite{kudus2023increased}. For radiology report cleaning, we removed all dates, numbers, punctuation marks, and any words or phrases containing genetic marker labels to avoid data leakage. 
   
Dataset 2 is different than Dataset 1 in terms of MRI scanner manufacturers and time range, which is 2018-2023, making it a proper option for independent model evaluation. 
For the course of this binary classification evaluation, we utilized 76 MR images from this dataset, without any accompanying radiology reports. Genetic marker labels were applied to measure the model's test performance, and segmentation masks served as ground truth for assessing the explainability.

\subsection*{Pretraining MRI-Report Contrastive Learning}
\label{CL}
At this stage, we train a CL framework on paired MR images and radiology reports based on the association between these modalities. The labels required for this task are 0 and 1, where 1 shows that an image-report pair is matched, while 0 indicates otherwise. 
To encode the MR images, we use 3D ResNet \cite{he2016deep}. It contains four 3D convolutional modules, batch-normalization, average-pooling, and dropout applied after each convolutional module except the last one. Additionally, we apply a self-attention layer to the last convolutional module of this model to make it focus on relevant parts of the image. Similar to the attention mechanism \cite{vaswani2017attention}, self-attention contains three main elements, namely query, key, and value, which can be obtained by transforming the output of the last convolutional module using three sets of trainable 3D convolutional weighs. Next, it calculates attention weights using the dot product between the transformed query and key. These attention weights are then  multiplied by the value, and the output can be added to the original image embedding in a residual manner to form a new weighted image embedding. 

As our dataset is relatively small, training the ResNet model from scratch can lead to overfitting. Therefore, we apply transfer learning by initializing the wights of this model with a set of pretrained weights, named ``Med3D" \cite{chen2019med3d} (MedicalNet), extracted by training 3D ResNet on a collection of 3D medical images including MRI. Since shallow layers in CNNs typically capture low-level features in contrary to deep layers that encode task-specific and high-level features, after loading the pretrained MedicalNet weights, we freeze the first two convolutional modules and fine-tune the last two ones. This can not only help stabilize the performance of the model but will also reduce the number of trainable parameters required by the model.

To convert the radiology reports into useful text representations, we feed them into a language transformer, Longformer \cite{Beltagy2020Longformer}. This transformer accepts input sequences of length up to 4096 tokens, making it suitable for encoding long texts such as MRI radiology reports. We initialize this model with Clinical Longformer \cite{li2022clinical} weights, which is a version of Longformer trained on MIMIC 3, a large dataset containing health-related information about patients admitted to critical care units \cite{johnson2016mimic}. Similar to the ResNet model, we freeze the first (ten) layers of Longformer and only fine-tune the last two ones.

\begin{figure*}[t]
\centering
\includegraphics[width=5.5in]{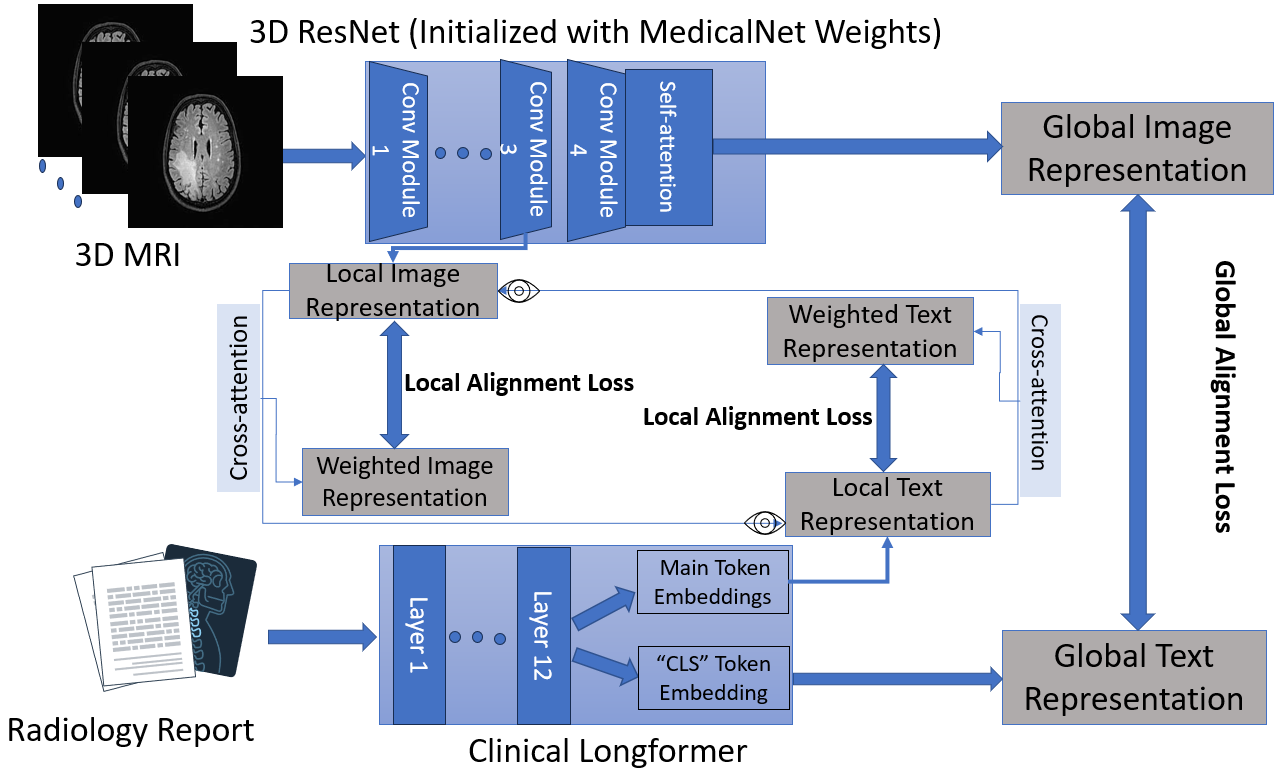} 
\caption{The Proposed MRI-Report Contrastive Learning Framework}
\label{cl}
\end{figure*}

Figure \ref{cl} demonstrates the overall framework of this MRI-report CL model. In the next subsections, we will discuss the loss function in detail, explaining each of the terms individually.

\subsubsection*{Global CL Loss Function}
\label{global}
The aim of our global CL loss term is to minimize the distance between similar global image and report embeddings, while maximizing that between dissimilar pairs. In other words, the global representations of the two modalities that correspond to the same datapoint, i.e., patient, should be highly close to each other in the embedding space, while the representations of unpaired images and reports should be far apart. In this way, the model can encode the two modalities in a way that would semantically match in the embedding space, resulting in cross-modal-aware representations. 

The global MRI representation is obtained by extracting the output of the self-attention layer, which is the weighted feature map of the last convolutional module with 512 channels. Furthermore, we consider the ``Classify Token (CLS)" embedding taken from the last attention layer in the Longformer transformer as the global text representation. CLS is a special token used by the transformer as an aggregation of the representation of the entire text, which has 768 channels and appears as the first element of the last layer's output \cite{Beltagy2020Longformer}. 

As shown by \cite{chen2020simple}, a multi-layer perceptron (MLP) head in a CL framework can effectively project the extracted representations into a space where both representations get the same shape and CL loss can be applied. We included this MLP head in our framework as well by feeding each global image and text representation into a sequence of a Linear layer, ReLU for non-linearity, and a second Linear layer. Hence, both representations maintain an equal number of 512 channels after passing through this layer.

For training this framework, we experimented with Triplet and Contrastive Losses and found that Triplet Loss leads to better downstream results (more details to be found in Appendix). Triplet loss \cite{schroff2015facenet} works by considering each global image/text representation in a batch as a reference called image/text ``anchor" and simultaneously minimizing the distance between this representation and the positive, i.e., corresponding, text/image representation as well as maximizing the distance between the anchor representation and a negative, i.e., disconnected, text/image representation. We utilize the cosine distance as our distance metric, which is computed as 1 minus the cosine similarity. 


We determine all three elements of the triplet loss, i.e., anchor, positive, and negative representations, based on data batches. Specifying the anchor image/text and the positive text/image representations is straight-forward as they are paired in the batches. For highlighting the negative representation, different methods can be applied. Random negative sampling, i.e., using a random representation from the batch which is disconnected with anchor is a potential method, but it may not lead to ideal representations. Hard negative sampling is an effective method that assigns a representation that has the lowest distance from the anchor, i.e., the most difficult one to be identified as negative, within a batch to the negative representation \cite{schroff2015facenet}, making the model learn useful representations from these difficult cases. In spite of effectiveness, this method increases the training time of the model as it requires computing 15 distance metrics for each anchor in our batches of size 16.

In this work, we propose a novel approach, which we call ``semi-hard negative sampling", to find the negative representations. Our approach works by sampling two random negative text/image representations for a certain image/text anchor in a batch and considering the one with a lower distance from the anchor as the negative representation. This method can be more effective than random sampling and also requires less training time than hard negative sampling by calculating only two distance metrics for each anchor in the batch.

\emph{Tumor location} has been identified as an important factor in different brain tumor diagnostic tasks \cite{fangusaro2024pediatric,trasolini2022mr}. Therefore, to improve our CL framework and representation learning process, we incorporate tumor location as an external variable into this framework. The idea behind it is that in cases where an MR image is not matched with a report, but they correspond to the same tumor location, the distance between their global representations in the embedding space should be less than that between the global representations of an unmatched MRI-report pair with different tumor locations.

To integrate tumor location into the global  loss, we add a new term. This term applies a coefficient in the range of (0,1) to the distance between the global representations of unrelated MRI-report pairs if they are associated with the same tumor location, pushing their distance to be shorter than that between the global representations of negative MRI-report pairs having different tumor locations. Equation \ref{eq1} indicates the mathematical description of our location-weighted global CL loss function.


\begin{equation}
\begin{aligned}
\label{eq1}
    \text{Global Loss} = \sum_{i=1}^{2}\sum_{j=1}^{n}(max(0,(1-\cos(a_{ij},p_{ij}))
    -(1-l_{j})(1-\cos(a_{ij},n_{ij})) 
    -c\times(l_{j})(1-\cos(a_{ij},n_{ij})) + m)) 
\end{aligned}
\end{equation}

 The above global loss operates based on the interactions between global MRI and report representations, where $i=1$ and $i=2$ demonstrate that the anchor is a global MRI and report representation, respectively. $n$ denotes the batch size, and $cos$ indicates Cosine Similarity. $a_{ij}$ demonstrates the $j^{th}$ global MRI/report representation in the batch as the anchor. $p_{ij}$ refers to the global report/MRI representation matched with the $j^{th}$ anchor, and $n_{ij}$ refers to a global report/MRI representation disconnected with this anchor. Moreover, $m$ (margin) is a hyperparameter that ensures a minimum distance between negative and positive representations, aiding in their distinct separation during the training process \cite{schroff2015facenet}, which is set to 0.25. $l_{j}$ denotes tumor location match, taking 1 if the location label of the $j^{th}$ MRI-report pair is the same and 0 otherwise. The location label is denoted by a unique integer assigned to each datapoint in the dataset, indicating the specific part of the brain where the corresponding tumor is located. This will be elaborated in the ``Datasets" subsection. Furthermore, $c$ is a coefficient in the range of (0,1), which is applied to the loss only if $l$ is equal to 1. The value of this coefficient is adjusted to 0.5 after hyperparameter tuning.

\subsubsection*{Local CL Loss Function}

The goal of adding this term to the CL loss function is to ensure that fine-grained image patches and report words corresponding to the same datapoint get encoded similarly. This is due to the fact that a certain local part of the report indicates the findings of radiologists regarding a certain patch of the associated MRI, and hence, they should be close to each other in the embedding space.

We assign the output of the third convolutional module in ResNet to  the local MRI representation as this layer typically captures fine-grained and low-level visual features. To obtain the local report embedding, similar to extracting the global one, we consider the output of the last layer in Clinical Longformer. As this transformer may break a word into several tokens to determine the attention scores based on its internal token dictionary, we first sum up the attention scores of multiple tokens corresponding to a single word based on the special token used for specifying these broken words. However, instead of taking the ``CLS" embedding, which was performed in the global CL loss, we use the embedding vector of all other tokens in the text as a single vector as the local text embedding.

To find the correlation between the extracted local image and text representations and feed them into the local loss, we utilize cross-attention, which is a specific type of the attention mechanism used for situations where the source and target of attention are from different modalities. To that end, we add two attention modules to our framework. The query of each of these modules comes from one of the image/text local embeddings, and the corresponding key and value are assigned to the embedding of the other modality. 
Each module returns two elements: cross-attention scores, associated with each of the key's and query's elements, and weighted query output, which pertains to a weighted representation of the query based on the cross-attention scores \cite{huang2021gloria,wang2022multi}.

Our local loss term consists of two sub-terms: A triplet loss between the local image representation and report-based-weighted local image representation, and another one between the local text representation and image-based-weighted local text representation. This local loss term is demonstrated in Equation \ref{eq2}. 

\begin{equation}
\begin{aligned}
\label{eq2}
    &\text{Local Loss} = \sum_{k=1}^{2}\sum_{i=1}^{2}\sum_{j=1}^{n}(max(0,(1-\cos(a_{ijk},p_{ijk}))-(1-\cos(a_{ij},n_{ijk})) 
    +m)) 
\end{aligned}
\end{equation}

The local loss function relies on the association between local MRI/report representations and their corresponding weighted counterparts, treating each modality as a distinct term. $k$ takes 1 or 2 if the term corresponds to local MRI and report representations, respectively. If the anchor represents an original local representation, then $i$ will be equal to 1; if the anchor represents a weighted local representation, then $i$ will be equal to 2. $j$, $n$, $cos$, and $m$ are defined similarly to the global loss (the previous subsection). $a_{kij}$ assigns the $j^{th}$ weighted/original local representation in the batch to the anchor. $p_{kij}$ shows the original/weighted local representation matched with the $j^{th}$ anchor, while $n_{kij}$ denotes an original/weighted local representation disconnected with this anchor.

\subsubsection*{Total CL Loss}
\label{tumor location}
 
Finally, our overall CL loss function is the summation of weighted global and local terms, as represented in Equation \ref{eq3}.

\begin{equation}
\fontsize{8.4}{8.1}\selectfont
\label{eq3}
     \text{Total CL Loss} = \alpha \times\text{Global Loss} + \beta \times \text{Local Loss} 
\end{equation}

Where $\alpha$ and $\beta$ are trainable parameters.

For training the whole CL architecture, which is developed in Pytorch, we apply AdamW \cite{loshchilov2017decoupled} optimizer with a learning rate of 0.0001. The batch size and number of epochs are set to 16 and 400, respectively. A complete list of hyperparameters can be found in Appendix.
All experiments are run on a ``Tesla V100-PCIE" GPU.


\subsection*{Downstream pLGG Genetic Marker Classification}
\label{DS}
\begin{figure}[ht!]
  \centering 
      \includegraphics[width=5.5in]{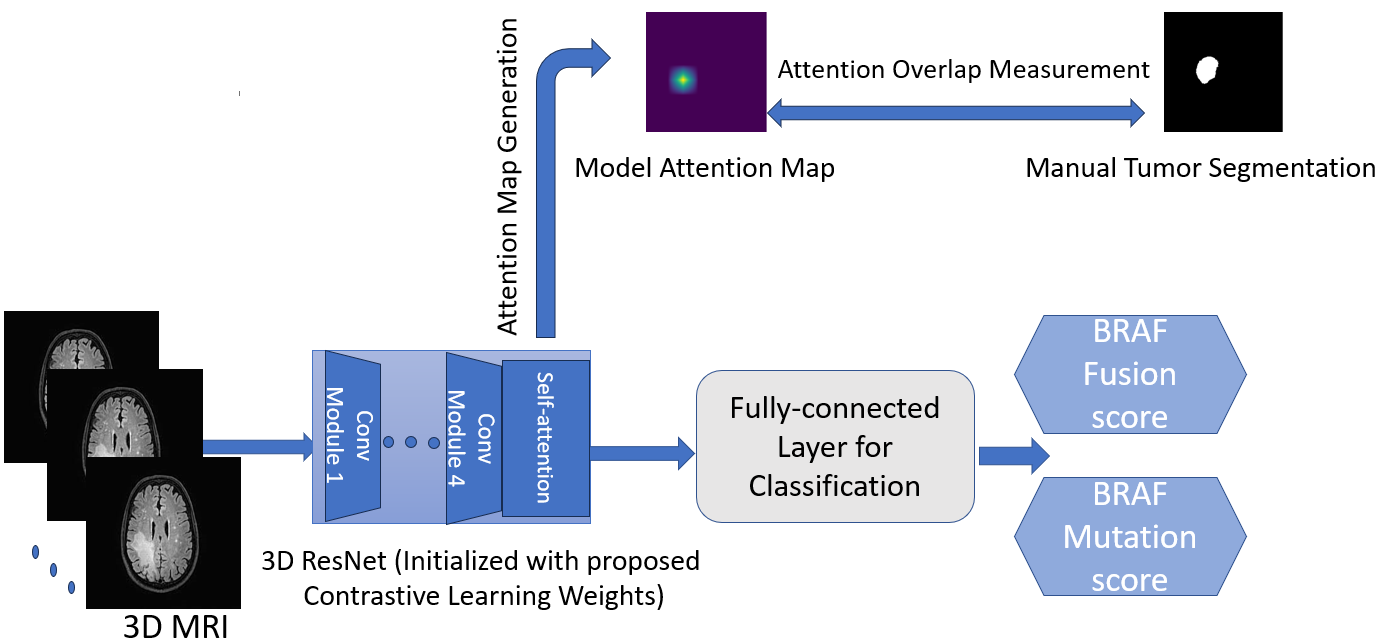} 
      \caption{The architecture of the downstream pLGG genetic marker classification model} 
      \label{ds} 
    \end{figure}

After training the CL framework, we apply only the learnt MRI representation to the downstream pLGG genetic marker classification task. In other words, we initialize a 3D ResNet model with the weights obtained from training the same model in our pretraining CL framework. Therefore, no report is required at this stage. Similar to the pretraining stage, we freeze the first two convolutional modules and fine-tune the last two ones. Additionally, we add a fully-connected layer on top of the convolutional modules to perform the classification. For training this network, we use Cross-entropy loss, AdamW optimizer with an initial learning rate of 0.0001, batch size of 16, and 20 epochs. StepLR scheduler is applied to adjust the learning rate by a factor of 0.5 every 10 epochs with a warm-up of 10 epochs. Figure \ref{ds} demonstrates the architecture used for this model. 

\subsection*{Evaluation Setting}

To measure the impact of our CL framework on the performance and explainability of the downstream classification task, we consider two baselines: a 3D ResNet model trained from scratch; and one initialized with MedicalNet weights and fine-tuned on the downstream classification labels. Subsequently, we compare the baseline models with the ResNet model pretrained using our CL framework and fine-tuned on the classification task.

We present the classification results based on the two aforementioned datasets. For internal evaluation of each of the five trained models in each experiment, we use a subset of Dataset 1 allocated for testing that model specified through cross-validation. External evaluation involves Dataset 2, where each of the five models is tested once in every experiment. Our metrics for each dataset are reported in a separate table, showing the mean and standard deviation (Std) (in parentheses) of the results obtained from testing each model on the respective dataset. 

\section*{Results}

\subsection*{Downstream Classification Performance Evaluation}
To compare the effect of each experiment on the classification performance of our downstream classification task, we use AUC, which represents the true positive rate versus false positive rate. Additionally, we employ Precision (Equation \ref{eq4} in Appendix), Recall (Equation \ref{eq5} in Appendix), and F1-score (Equation \ref{eq6} in Appendix) as secondary metrics. Table \ref{table2} indicates these metrics when applied to Dataset 1, while Table \ref{table3} demonstrates the external validation of the classification performance on Dataset 2.

\begin{table*}[t]
  \centering 
  \caption{Mean Test Classification performance results of 3D ResNet on Dataset 1} 
  
  \begin{tabular}{lllll}
  \toprule
    \textbf{Model Initialization} & \textbf{AUC} & \textbf{Precision}  & \textbf{Recall}  & \textbf{F1-score}\\
    \midrule
    Randomly & 0.748 ($\pm0.083$) & 0.400 ($\pm0.242$) & 0.552 ($\pm  0.298$) & 0.441 ($  \pm0.232$)\\ 
    With MedicalNet  & 0.790 ($\pm0.101$) & 0.380 ($\pm0.064$) & \textbf{0.918} ($\pm0.086$) & 0.534 ($\pm0.068$)\\ 
    With proposed CL &  \textbf{0.877} ($ \pm0.072$) &  \textbf{0.708} ($\pm0.189$) &  0.638 ($\pm0.161$) & \textbf{0.652} ($\pm0.137$) \\ 
    \bottomrule
  \end{tabular}
  \label{table2} 
\end{table*}

In internal validation (Table \ref{table2}), as it can be observed, transfer learning, i.e., loading pretrained weights, is highly helpful in this classification task as the model trained from scratch shows 
 the poorest performance among the three experiments. In contrast, the ResNet model initialized with the proposed CL weights outperforms the two baselines with a higher AUC (0.877), precision (0.708), and f1-score (0.652). However, it is worth noting that initializing the model with MedicalNet weights surpasses the proposed model solely in terms of recall (0.918 versus 0.638).

 \begin{table*}[t]

  \centering 
  \caption{Mean External classification performance results of 3D ResNet on Dataset 2} 
  \begin{tabular}{lllll}
  \toprule
    \textbf{Model Initialization} & \textbf{AUC} & \textbf{Precision}  & \textbf{Recall}  & \textbf{F1-score}\\
    \midrule
    Randomly & 0.467 ($\pm0.115$) & 0  & 0  & 0 \\ 
    With MedicalNet & 0.418 ($\pm0.062$) & 0.069 ($\pm0.138$) & 0.061 ($\pm0.122$) & 0.065 ($\pm 0.130$)\\ 
    With proposed CL & \textbf{0.757} ($\pm0.033$) & \textbf{0.599} ($\pm0.089$) & \textbf{0.524} ($\pm0.152$) & \textbf{0.534} ($\pm0.058$)\\ 
    \bottomrule
  \end{tabular}
  \label{table3} 
\end{table*}

In external validation (Table \ref{table3}), both baselines exhibit  failure and achieve very low classification metrics. However, loading the proposed CL weights into the 3D ResNet model significantly enhances its performance across all metrics. In particular, it improves the AUC of the randomly-initialized model and the one initialized with MedicalNet weights by 62.1\% and 81.1\%, respectively. The results of t-tests showing the significance of these improvements are provided in Appendix.

\subsection*{Downstream Classification Explainability Evaluation}

We assess the explainability of the classification model by first visualizing its attention heatmaps based on the weights extracted from the self-attention module. An attention heatmap indicates the parts of an image attended by the model for making a certain prediction. Subsequently, we normalize the heatmaps to the range of [0,1] and use a threshold of 0.01, determined through visual examination of the heatmaps, for binarization. Next, we compare these attention maps with the manual tumor segmentation masks as the explainability ground-truth. Higher overlap between them indicates that the model has focused on relative parts of the image where the biopsy was taken for determining the genetic marker of the tumor, and therefore, achieves higher explainability. We apply 2D and 3D Dice scores (Equation \ref{eq7} in Appendix) to quantitatively measure this overlap. 3D Dice is based on the whole MRI volume, while 2D Dice is based on the MRI slice that  has the largest cross-section with the tumor segmentation. 

\begin{figure}[h] 
\centering
\begin{tabular}{c}
\vspace{-5mm}
\includegraphics[width=0.5\textwidth]{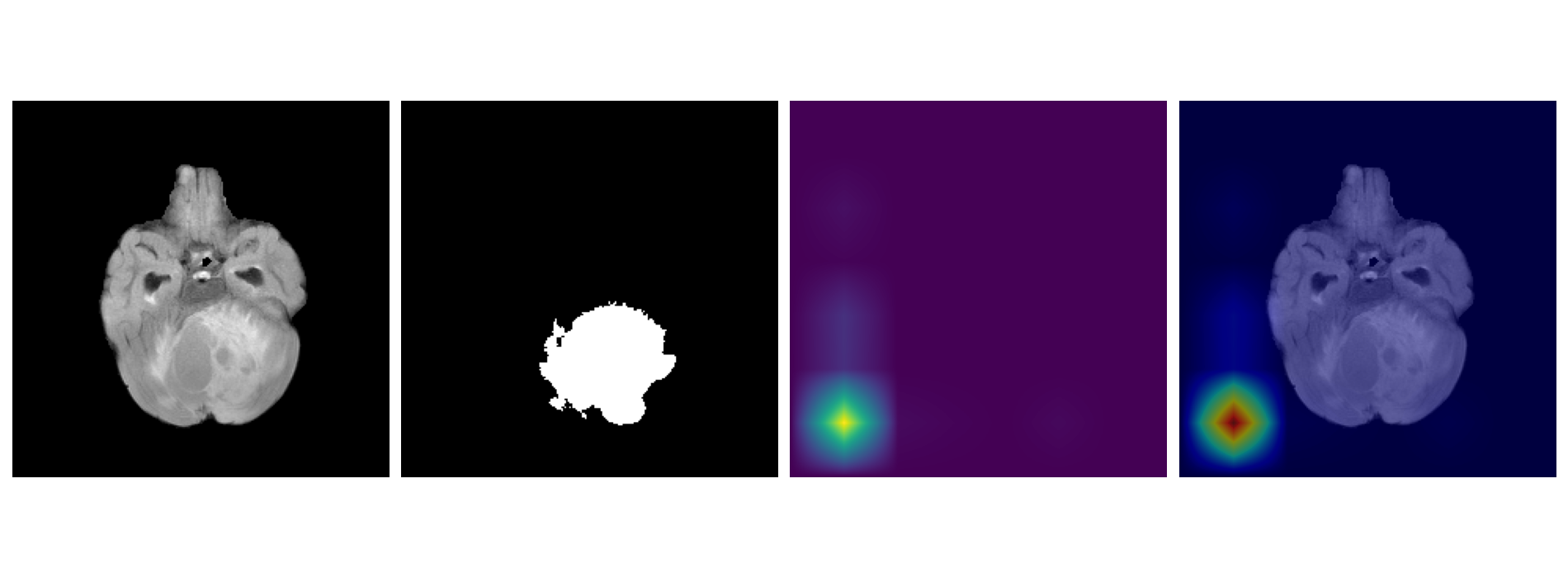}\\
\vspace{-1mm}
(a)\\
\vspace{-5mm}
\includegraphics[width=0.5\textwidth]{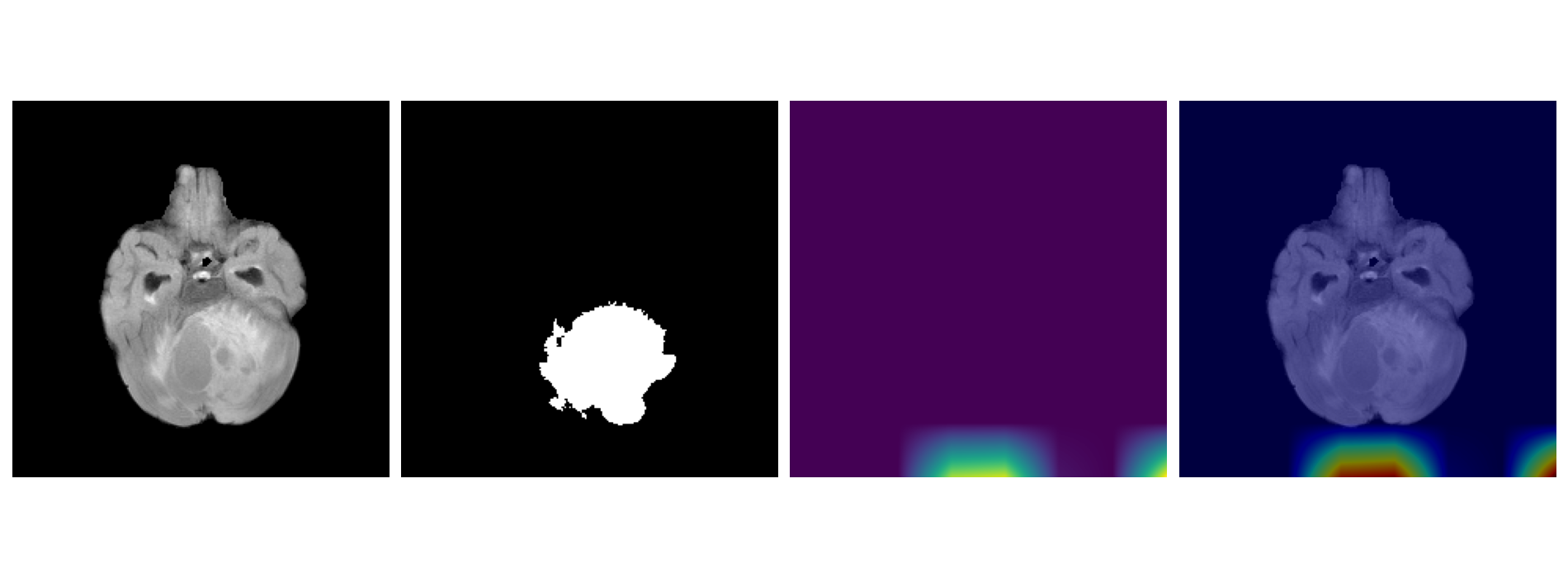}\\
\vspace{-0.9mm}
(b)\\
\vspace{-5mm}
\includegraphics[width=0.5\textwidth]{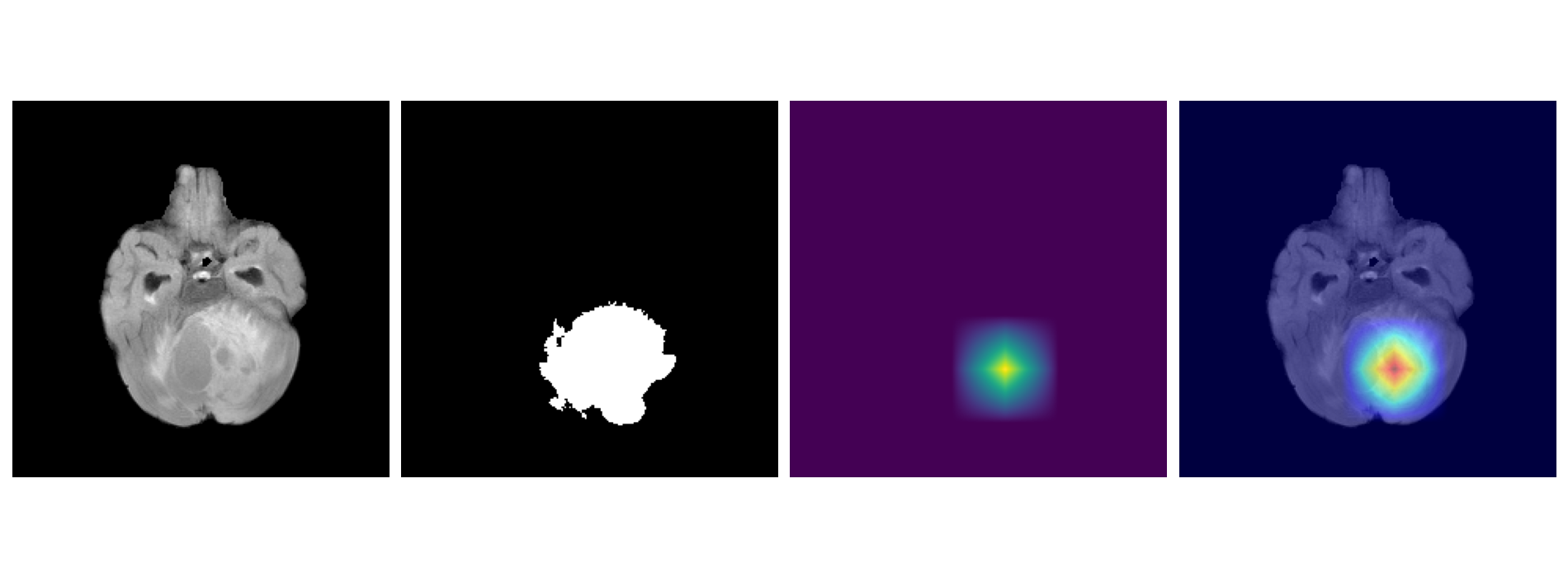}\\
\vspace{-0.9mm}
(c)\\
\end{tabular}
\caption{From left to right: MRI slice, manual tumor segmentation mask slice, model attention map slice, and MRI with overlaid attention map, for a sample in Dataset 2, when 3D ResNet is initialized: (a) randomly, (b) with MedicalNet weights, and (c) with proposed CL weights}
\label{result}
\end{figure}

Figure \ref{result} depicts a slice of MRI and the corresponding tumor segmentation mask with the largest cross-section with the tumor for a datapoint in Dataset 2, along with the associated model attention heatmaps extracted from the three classification experiments. Tables \ref{table4} and \ref{table5} demonstrate the quantitative measurements of the attention overlap between the segmentation masks and model heatmaps based on the first and second datasets, respectively.

\begin{table}[ht]

  \centering 
  \caption{Mean Test classification explainability  results of 3D ResNet on Dataset 1} 
  \begin{tabular}{lll}
  \toprule
    \textbf{Model Initialization} & \textbf{2D Dice Score} & \textbf{3D Dice Score}\\
    \midrule
    Randomly & 6.1\% 
 ($\pm8.5\%$)  &  2.0\% ($\pm2.8\%$)\\ 
    With MedicalNet & 22.5\% ($\pm10.9\%$)& 8.9\% ($\pm4.8\%$) \\ 
    With proposed CL & \textbf{31.1\%} ($\pm2.4\%$) & \textbf{15.8\%} ($\pm2.1\%$)\\ 
    \bottomrule
  \end{tabular}
  \label{table4} 
\end{table}
\begin{table}[ht!]

  \centering 
  \caption{Mean External Classification Explainability results of 3D ResNet on Dataset 2} 
  \begin{tabular}{lll}
  \toprule
    \textbf{Model Initialization} & \textbf{2D Dice Score} & \textbf{3D Dice Score}\\
    \midrule
     Randomly & 4.4\% ($\pm4.2\%$)  &  1.8\% ($\pm1.9\%$)\\ 
    With MedicalNet & 4.5\% ($\pm4.6\%$) & 1.5\% ($\pm1.7\%$)\\ 
    With  proposed CL & \textbf{30.7}\% ($\pm0.3\%$) & \textbf{16.0}\% ($\pm0.3\%$)\\ 
    \bottomrule
  \end{tabular}
  \label{table5} 
\end{table}

According to Table \ref{table4}, the baseline models achieve very low explainability on Dataset 1, as indicated by low Dice scores. Nonetheless, loading the proposed CL weights into the 3D ResNet model yields superior explainability, with 2D and 3D Dice scores of 33.1\% and 15.8\%, respectively. The external validation of the explainability on Dataset 2 exhibits a similar trend, and the proposed model outperforms the baselines based on both 2D and 3D Dice scores. This is also demonstrated by comparing the attention heatmaps of this model with that of the baselines in Figure \ref{result}, where the heatmap of the CL-based-initialized classification model represents higher overlap with the manual segmentation.






\subsection*{Ablation Studies}

To show the effectiveness of the proposed CL framework, we conducted five ablation experiments by changing its different components. The first three experiments pertain to training the framework using the global loss term only, local loss term only, and global and local terms without integrating tumor location. Furthermore, in Experiment 4, we replaced Triplet Loss with Contrastive Loss. Finally, in the last experiment, we employed hard negative sampling, i.e., selecting the datapoint closest to an anchor in a batch as the negative sample.
Table \ref{tab66} demonstrates the results of applying the downstream pLGG genetic marker classification task initialized with the weights of each of these models to our external dataset. 
\begin{table}[ht!]

  \centering 
  \caption{The Mean Classification Results of the Ablation Experiments on Dataset 2}
  \begin{tabular}{lll}
  
  \toprule
    \textbf{Experiment} & \textbf{Mean AUC} & 
    \textbf{Mean F1-score}\\
    \midrule
    Global only  & 0.608 & 0.124 \\ 
    Local only & 0.567 & 0.434 \\ 
    Tumor location-excluded & 0.683 & 0.523 \\
    Contrastive loss& 0.692 & \textbf{0.559} \\
    Hard negative sampling & 0.450 & 0.364 \\
    Proposed method & \textbf{0.757} &  0.534 \\
    \bottomrule
  \end{tabular}
  \label{tab66} 
\end{table}

As we can observe, based on AUC, our primary performance metric, our model's performance is notably higher than that of the aforementioned experiments, highlighting the significant impact of the various components we have integrated into it. Nonetheless, the model trained with Contrastive Loss surpasses the proposed architecture based on F1-score.

Regarding the loss function, the superior performance of Triplet Loss over Contrastive Loss in terms of AUC could be potentially due to the small size of our dataset, where incorporating all available negative samples in a batch into the loss leads to overfitting. Consequently, given the limited size of the data, using Triplet loss by determining a suitable negative representation and simultaneously including that in the loss along with the positive representation corresponding to an anchor can efficiently encode generalized patterns in the data and avoid overfitting. In contrast, other CL-related works typically utilize thousands of datapoints, enabling much larger batch sizes, and including all negative representations in a batch in gradient updates helps the model learn diverse patterns available in the data. Therefore, we hypothesize that in low data regimes, Triplet Loss can outperform other CL-related loss functions and lead to higher performance in downstream tasks.

Moreover, due to the importance of selecting a negative sample that significantly helps regulate the distance between mismatched image and text representations, we tried semi-hard (random negative sampling sizes of 2 and 8) and hard negative sampling (negative sampling size of 15). In other words, in these experiments, we randomly chose 2, 8, and 15 negative sample representations within a batch of size 16, and selected the negative representation having the lowest distance from the anchor representation. Using a random subsampling size of 2 led to the best performance in the downstream classification task. When setting this variable to 8 or 15, we found that in many batches, the model would end up selecting the same representation as the hard negative sample for most of the datapoints in the batch. This can be potentially due to the small dataset size and the resulting small batch size, leading to a lack of diverse representations within each batch, and the appearance of outliers. 


\section*{Discussion}

In this study, we developed a CL-based framework on brain MR images and radiology reports to learn the association between the global MRI and radiology report representations as well as the local MRI patches and radiology report words. Our framework also incorporates tumor location as an external variable for enhancing the representation learning procedure. For training this network, we had access to only 341 image-report pairs, notably smaller than other proposed CL-based frameworks utilizing thousands or millions of datapoints \cite{chen2020simple, huang2021gloria,wang2022multi}. To overcome this challenge and avoid overfitting, we reduced the number of trainable parameters by freezing the first layers in both image and text encoders as well as applying regularization techniques, e.g., dropout. 

Despite the small size of the dataset, we demonstrated the effectiveness of the learnt MRI representations in enhancing the performance and explainability of the pLGG genetic marker classification as a downstream task. 
We indicated that initializing a 3D ResNet model with the weights learnt from the proposed CL-based framework significantly improves the performance of the same model trained from scratch or initialized with MedicalNet weights on both datasets, achieving AUC of 0.877 and 0.757 on Datasets 1 and 2, respectively. Moreover, loading the pretrained weights into this model increases the overlap between the model's attention maps and radiologist-specified tumor segmentation masks to 31.1\% on Dataset 1 and 30.7\% on Dataset 2, i.e., boosting the model's explainability. This underscores the crucial role of radiology reports in directing the MRI-based model's attention towards relevant image areas.

These findings, specifically the external results, demonstrate that our framework is capable of dealing with various data shifts, including those caused by chronological and scanner variations. Notably, the attention maps of the proposed model have significantly higher overlap with the ground-truth tumor segmentation masks compared to the baselines. Moreover, this model performs much better on the external dataset, as indicated by the provided performance metrics. The baseline models, on the other hand, are completely unable to generalize to this dataset, where they rely on out-tumor regions for making a diagnosis.

From a clinical standpoint, our proposed CL-based framework can be applied to enhance the performance of different downstream diagnostic tasks, one of which was explored in this work. Improving the classification of pLGG genetic markers can aid in non-invasive tumor prognosis and efficient targeted treatment planning, eliminating the need for biopsy. Furthermore, the boosted explainability of this task can build trust among radiologists and facilitate the integration of these models into the clinical workflow as a helping hand to radiologists. Ultimately, this could significantly alleviate radiologists' workload and heighten their efficiency. 

From the technical perspective, this framework can improve the generalizability of DL models using the correspondence between MR images and radiology reports, without requiring any human-provided labels. This is particularly important in the healthcare domain, where data annotation is a labor-intensive and challenging task. Fine-tuning this network on potential downstream tasks can significantly reduce the required amount of labelled data and improve the performance. Moreover, the local alignment module of this framework can learn fine-grained associations between imaging features and textual report components, guiding the attention of the image-based model to pertinent regions, thereby leading to higher explainability in downstream tasks.

In spite of showing the effectiveness of the proposed CL framework in improving the explainability and performance of pLGG genetic marker classification, there are a number of limitations in this work. First, our dataset is small, potentially affecting the generalizability of the framework, as shown by relatively low external validation results. As a next step, we will use data augmentation techniques to increase the number of datapoints used in the CL architecture to enhance both internal and external test performance. Second, we investigated the performance and explainability of pLGG genetic marker classification and did not consider downstream tasks that require no fine-tuning. In future work, we will explore other tasks, such as zero or few-shot classification, to better demonstrate the efficiency of the proposed framework. 

\section*{Conclusion}

In this work, we introduced an MRI-report CL architecture based on the global and local associations between these modalities and leveraged tumor location to enhance the learnt representations. Applying the proposed network to the downstream pLGG genetic marker classification, we observed an AUC of 87.7\%, 2D Dice score of 31.1\%, and 3D Dice score of 15.8\%, which are significantly higher than the baseline results. 
Thus, this framework not only reduces the amount of annotated data required by DL models for downstream imaging-based diagnostic tasks but can also raise the efficiency of radiologists in performing such tasks and lead to better patient outcomes.

\section*{Data Availability Statement}

The datasets generated and/or analyzed during the current study are available from the corresponding author
on reasonable request pending the approval of the institution(s) and trial/study investigators who contributed
to the dataset.

\bibliography{main.bib}


\section*{Figure Legends}

\textbf{Figure 1.} The Proposed MRI-Report Contrastive Learning Framework

\noindent \textbf{Figure 2.} The architecture of the downstream pLGG genetic marker classification model

\noindent \textbf{Figure 3.} From left to right: MRI slice, manual tumor segmentation mask slice, model attention map slice, and MRI with
overlaid attention map, for a sample in Dataset 2, when 3D ResNet is initialized: (a) randomly, (b) with MedicalNet weights,
and (c) with proposed CL weights

\section*{Acknowledgements}

This research was supported by Natural Sciences and
Engineering Research Council of Canada (NSERC).

\section*{Author contributions statement}

S.K. and F.K. contributed to the design of the concept and study. S.K. contributed to the implementation of machine learning modules and running the experiments. S.K. contributed to preprocessing
of textual data. U.T., and C.H. contributed to collecting, reviewing, and providing clinical data and genetic markers. M.W. and B.E.W. contributed to reviewing the imaging data and providing tumor segmentation
for the data. S.K. wrote the first draft of the manuscript and all authors contributed to the reviewing and editing
of the manuscript. All authors read and approved the final manuscript.

\section*{Additional information}

The authors declare no competing interests.

Correspondence and requests for materials should be addressed to F.K.

The corresponding author is responsible for submitting a \href{http://www.nature.com/srep/policies/index.html#competing}{competing interests statement} on behalf of all authors of the paper. This statement must be included in the submitted article file.


\newpage
\appendix
\section*{Appendix}
\label{appendix}
\subsection*{Dataset Distribution}

Table \ref{tab5} indicates the distribution of the genetic marker labels in our studied datasets.

\begin{table}[b]

  \centering 
  \caption{The distribution of the genetic marker classes in Dataset 1 (used for classification fine-tuning) and Dataset 2 (used for independent validation)}
  \begin{tabular}{llll}
  \toprule
    \textbf{Dataset} & \textbf{BRAF Fusion} & \textbf{BRAF Mutation}  & \textbf{Other} \\
    \midrule
    Dataset 1 & 137 & 67 & 137 \\ 
    Dataset 2 & 48 & 28 & 54 \\ 
    \bottomrule
  \end{tabular}
  \label{tab5} 
\end{table}

\subsection*{Dataset Choice}

We used an internal dataset for developing our experiments instead of open-source datasets for two main reasons:

\begin{itemize}
    \item Training the proposed CL framework required a dataset containing MR images, radiology reports and tumor location information. We did not find any open-source datasets that included all of these elements.
    \item For fine-tuning the CL framework on the plGG genetic marker classification task, we needed labels specific to a pediatric population. The used dataset is one of the richest available in terms of providing such labelled images.
\end{itemize}

\subsection*{Evaluation Metric Details}
\vspace{0.5cm}
\begin{equation}
\label{eq4}
     \text{Precision} = \frac{TP}{TP+FP}
\end{equation}

\begin{equation}
\label{eq5}
     \text{Recall} = \frac{TP}{TP+FN}
\end{equation}

Where $TP$ (true positive), $FP$ (false positive), and $FN$ (false negative) demonstrate the counts of BRAF V600E Mutation cases correctly classified as BRAF V600E Mutation, BRAF Fusion cases incorrectly classified as BRAF V600E Mutation, and BRAF V600E Mutation cases incorrectly classified as BRAF Fusion, respectively.

\begin{equation}
\label{eq6}
     \text{F1-score} = \frac{2\times Precision\times Recall}{Precision+Recall}
\end{equation}

\begin{equation}
\label{eq7}
     \text{Dice\:Score} = \frac{2\times Overlap\:Area}{Total\:Area}
\end{equation}

Here, $Overlap\:Area$ refers to the intersection of two images, while $Total\:Area$ represents the combined area encompassed by the images.

\subsection*{Number of Trainable Parameters}

Table \ref{param} indicates the number of trainable parameters used in our pretraining contrastive learning and downstream classification frameworks:

\begin{table}[h]
    \centering
\begin{tabular}{|c|c|c|c|c|c|}

  \hline
  \textbf{Framework} & \textbf{Image Encoder}  & \textbf{Text Encoder} & \textbf{Cross-attention} & \textbf{MLP Heads} & \textbf{Total} \\
  \hline
  Pretraining contrastive learning & 31759873  & 18900736 & 789504 & 525312 & 52107009 \\
  \hline
  Downstream classification & 31759873  & N/A & N/A & 131585 & 31891458 \\
  \hline
  
\end{tabular}
\caption{Number of Trainable Parameters used in Our Models}
    \label{param}
\end{table}

\subsection*{Hyperparameters}

Tables \ref{CL_hyper} and \ref{ds_hyper} demonstrate the hyperparameters used in our pretraining CL and downstream classification task, respectively. The optial value for each hyperparameter was selected based on validation AUC.

\begin{table}[ht!]
    \centering
\begin{tabular}{|c|c|c|c|c|c|}

  \hline
  \textbf{Hyperparameter} & Dropout rate  & Number of epochs & Learning rate & Negative sampling size in Triplet Loss & Margin \\
  \hline
  \textbf{Tuned Values} & 0.25,0.5  & 230,340,500 & 8e-5,1e-4,5e-4 & 2,8,15 & 0.25,0.4,1 \\
  \hline
  \textbf{Optimal Value} & 0.25  & 340 & 1e-4 & 2 & 0.25 \\
  \hline
  
\end{tabular}
\caption{Tuned Hyperparameters in Our CL Framework}
    \label{CL_hyper}
\end{table}

\begin{table}[h]
    \centering
\begin{tabular}{|c|c|c|c|}

  \hline
  \textbf{Hyperparameter} & Number of epochs & Learning rate & dropout rate \\
  \hline
  \textbf{Tuned Values} & 10,15,20  & 1e-4,1e-3,1e-2 & 0.1,0.2,0.25,0.5  \\
  \hline
  \textbf{Optimal Value} & 20 & 1e-4& 0.25 \\
  \hline
  
\end{tabular}
\caption{Tuned Hyperparameters in Our Downstream Classification Architecture}
    \label{ds_hyper}
\end{table}

\subsection*{Number of Frozen Layers}

Since deep layers in convolutional neural networks typically capture high-level features and shallow layers extract low-level and task-agnostic features, we decided to freeze initial layers and fine-tune the last ones (along with fully-connected modules in the downstream classification task). Due to the computational resources and small dataset size, we had to fine-tune at most two layers in each of the image and text encoders to avoid overfitting. We experimented with fine-tuning one or two layers in each encoder and found that fine-tuning the last two layers in both encoders achieves the best performance in the downstream classification task.

\subsection*{Statistical Tests}

Tables \ref{t_test} and \ref{t_test_2} show the results of Paired  T-test for evaluating the significance of the external downstream classification and explainability results on Dataset 2.

As highlighted in Table \ref{t_test}, the extracted p-values for both classification metrics, i.e., AUC and F1-score, are below 0.05, indicating that the improvements in external classification results over both baselines are statistically significant. Similarly, according to Table \ref{t_test_2}, the p-values for both explainability metrics, 2D and 3D Dice scores, are less than 0.05 across both experiment pairs.

\begin{table}

  \centering 
  \caption{The Paired T-test Results for the External Classification Metrics  on Dataset 2}
  \begin{tabular}{lll}
  
  \toprule
    \textbf{Experiment Pairs} & \textbf{P-value for AUC} & 
    \textbf{P-value for F1-score}\\
    \midrule
    initialized with proposed weights, initialized randomly  & 0.0036 & 0.0000\\ 
    initialized with proposed weights, initialized with MedicalNet & 0.0000 & 0.0005 \\ 
    
    \bottomrule
  \end{tabular}
  \label{t_test} 
\end{table}

\begin{table}

  \centering 
  \caption{The Paired T-test Results for the External Explainability Metrics on Dataset 2}
  \begin{tabular}{lll}
  
  \toprule
    \textbf{Experiment Pairs} & \textbf{P-value for 2D Dice} & 
    \textbf{P-value for 3D Dice}\\
    \midrule
    initialized with proposed weights, initialized randomly  & 0.0001 &  0.0001\\ 
    initialized with proposed weights, initialized with MedicalNet & 0.0002 & 0.0000 \\ 
    
    \bottomrule
  \end{tabular}
  \label{t_test_2} 
\end{table}

\end{document}